# Angle Diversity Trasmitter For High Speed Data Center Uplink Communications

Abrar S. Alhazmi[1], Sanaa H. Mohamed[1], Osama Z. Alsulami[1], T. E. H. El-Gorashi[1], and Jaafar M. H. Elmirghani[1]
[1]School of Electronic and Electrical Engineering, University of Leeds, LS2 9JT, United Kingdom
elasal@leeds.ac.uk, s.h.h.mohamed@leeds.ac.uk, ml15ozma@leeds.ac.uk, t.e.h.elgorashi@leeds.ac.uk,
j.m.h.elmirghani@leeds.ac.uk

*Abstract*—This paper proposes an uplink optical wireless communication (OWC) link design that can be used by data centers to support communication in spine and leaf architectures between the top of rack leaf switches and large spine switches whose access points are mounted in the ceiling. The use of optical wireless links reduces cabling and allows easy reconfigurability for example when data centres expand. We consider three racks in a data center where each rack contains an Angle Diversity Transmitter (ADT) positioned on the top of the rack to realize the uplink function of a "top-of-the-rack (ToR)" or a "leaf" switch. Four receivers are considered to be installed on the ceiling where each is connected to a spine switch. Two types of optical receivers are studied which are a Wide Field-of-View Receiver (WFOVR) and an Angle Diversity Receiver (ADR). The performance of the proposed system is evaluated when the links run at data rates higher than 19 Gbps. The results indicate that the proposed approach achieves excellent performance using simple On-Off Keying (OOK) modulation.

*Keywords— Optical Wireless Communication (OWC), Angle Diversity Transmitter (ADT), Angle Diversity Receiver (ADR), Data centers, Uplink design, Top-of-the-Rack (ToR), Spine and leaf data centers.*

## I. INTRODUCTION

Data centers are essential in the future data centric economy [1]. Due to the recent outbreak of COVID-19, several organizations had to completely transition to cloud platforms. Hence, local and private data centers with limited-access have undergone revolutionary changes and have begun to adopt cloud solutions with the assistance of various channels, such as private or public cloud providers and colocation. The main trend observed during the outbreak of COVID-19 involved a widespread transition to digitalization and the extensive use of online platforms and virtualization due to quarantine measures adopted by local and federal governments. Additionally, with the introduction of the sixth-generation (6G) of communication systems with hyper-connectivity features, communication frameworks must be supported by high data rates in access networks [2]. In the near future, Optical Wireless Communication (OWC) systems will have the potential to enable the considerably high data rate requirements in 6G, as it continues to be a rapidly evolving field [3].

Access networks traffic has increased along with end users' demand for data, and higher levels of service demand that must be fulfilled by data centers. Nonetheless, the network architecture of conventional data centers cannot perform the role of a multi-tenant next-generation data center [1]. Data centers are also facing other challenges, such as maintenance, performance, reliability, and scalability. Therefore, there is a dire need for easily deployable and scalable data centers that can fulfil the requirements of dynamic tenant applications while improving management flexibility, curtailing infrastructure costs, and reducing the energy consumption [4] [5, 6].

Traditional fiber/cable connections in data center networks suffer from flexibility, scalability, and bandwidth overload issues [1]. One potential and promising solution for data center networking is OWC systems that are capable of meeting the demands of future 6G communication systems while offering high scalability and flexibility, low power consumption and high data rates [4, 5, 7]. The integration of wireless connections in data centers significantly improves the operational and maintenance efficiency by reducing the need for cabling and potentially cooling equipment [8], [9], [10].

In OWC systems, the data can be exchanged between transceivers through Infrared (IR) or Visible Light (VL) [11]. Light Emitting Diodes (LEDs) or Laser Diodes (LDs) can be used to transfer light from a transmitter to an optical receiver through free-space and can then be integrated with fiber optic systems [12]. Compared to Radio Frequency (RF) transmission, additional advantages can be achieved through IR [13-15], including large transmission bandwidths and immunity against interference caused by adjacent electrical devices. Furthermore, IR systems have relatively short transmission ranges and high data rates [16], so they are ideal for indoor settings and appropriate for data centers that require high data rates. According to the results of various studies, data transfer rates of up to 25 Gb/s and beyond can be transmitted per link by OWC frameworks within indoor settings [17-26]. In addition, experts have endorsed the commercial use of OWC systems, provided the cost of both the receiver and the transmitter components become affordable, followed by a surge in data requirements [27-30].

This paper proposes a novel method to enable uplink communication in Spine and leaf data centers that contain a number of racks each with an OWC-based leaf switch and one layer of spine switches in the ceiling. The proposed design architecture utilizes Angle Diversity Transmitters (ADTs) where three racks are considered here. The main advantage of using ADTs for data centers is that they provide multiple beams and hence achieve spatial multiplexing which improves the communication performance in the data center by using diversity approaches [31-35]. Two different types of receivers are considered: an Angle Diversity Receiver (ADR) and a Wide Field-of-View Receiver (WFOVR). The ADR uses different branches, each with small field-of-view, to gather signals, minimize the interference that can come from reflected beams and mitigate the background noise. On the other hand, the WFOVR collects higher optical signal power compared to ADR, but it also collects more noise and interference power. The data rates when using each of these receivers are then compared.

The remainder of this paper is organized as follows. Section II discusses some related work. Section III presents the system model including the data center setup and the optical receiver and transmitter designs. Section IV describes the simulation environment and discusses the corresponding results. Finally, Section V concludes the paper.

## II. RELATED WORK

Previous studies have proposed improvements in the design of data centers using IR and Visible Light Communication (VLC) systems for uplink and downlink communication, to achieve high data rates. In [36], the authors indicated that in VLC systems, one of the key challenges is the provision of high data rates. For downlink communication, the authors proposed a data center design employing Red, Yellow, Green, and Blue (RYGB) Laser Diodes (LDs), which were used as transmitters. Delay spread and Signal-to-Noise Ratio (SNR) were investigated using three types of receivers: a WFOVR, a 50-pixel imaging receiver (ImR), and a 3-branch ADR and data rates of up to 14.2 Gbps were achieved using the proposed systems. In [37], the researchers designed an optical wireless downlink for a data center integrated with Wavelength Division Multiple Access (WDMA). High modulation bandwidths were achieved using transmitters powered by RYGB LDs. For multiple racks, synchronized communication from the same light unit was provided using a WDMA scheme based on the RYGB LDs.

In [38], the researchers proposed various applications driven exclusively by high data rates, which can be achieved via IR uplinks. In the proposed architecture, three data center racks were deployed, each equipped with its own ADT transmitter. Four WFOVRs were fixed to the data center's ceiling, and the input to a spine switch was integrated with each of these receivers. While each link operated at a data rate above 7.14 GB/s, the proposed system's performance was assessed using the space or wavelength dimensions and higher data rates were achieved through multiple links. The proposed system was capable of achieving high data rate while using simple On Off Keying (OOK). In this paper, we extend the work in [26] by Using an ADR to minimize the interference in the data center attributed to the reflections and by considering the first-order and second-order reflections in the evaluation of the received power.

The authors in [39] and [40] reported on an OWC interconnection network solution for data centers, FireFly. In this approach, all connections are wireless and reconfigurable, and the ToR switches are removed. By creating a proof-of-concept prototype of a compact steerable form factor FSO device, the authors established the practicality of this design and constructed the algorithm necessary for the network management.

## III. SYSTEM MODEL

### A. Data Center Structure

We assume that the data center measures 8 m × 8 m × 3 m (length × width × height) in size [29], [38]. A vacant, unfurnished room without any windows or doors is used. In this work, the data center has three racks. The rack dimensions considered are 1.6 m × 1.2 m × 1.75 m (length × width × height, and each rack has a leaf (i.e. a ToR) switch as shown in Figs. 1 and 2. There are four equally-spaced spine switches whose inputs are attached to the room's ceiling as can be seen in Figs. 1 and 2. An ADT transmitter with broad beams and an ADR are located at the top of each rack and each branch directs its Line-of-Sight (LoS) link at one of the receivers.

Reflections from the room's four walls, ceiling, and floor are modeled using a ray-tracing algorithm [41]. The surface of each of the room's walls is divided into smaller zones with area $dA$ with a reflection coefficient of $\rho$. According to [42], plaster walls reflect light beams in a Lambertian pattern. In this work, we use a reflection coefficient of 0.8 for the ceiling and the walls and 0.3 for the floor [10]. Each element on each surface acts as an emitter reflecting the received beam as a Lambertian pattern with an emission order of $n$ equal to 1. In this study, a zone with a surface area of 5 cm × 5 cm is selected as the surface component for first-order reflections and 20 cm × 20 cm for second-order reflections, which keeps the calculation time of the ray tracing within a feasible range [29]. Note that the third and fourth order reflections are not taken into consideration because the impact will be negligible [14], [41], [42]. The Communication Floor (CF) is set to be 0.25 m above the main floor. All communications involved occur above the CF. The ray-tracing algorithm in [41] is utilized to conduct the simulations.

### B. Optical Transmitter Design

Similar to the work in [20], [23], [27], [43], this study adopts the concept of ADT within a data center, as shown in Figs. 1 and 2. Three ADT transmitters are located at the top of the racks similar to [26] which are in three locations within the data center. More specifically, the first ADT transmitter is located at the corner with coordinates (1.3m, 1.6m, 2m), the second transmitter is located at the center of the room with coordinates (4m, 4m, 2m), and the third transmitter is located at the room corner with coordinates (4m, 6.7m, 2m).

The locations of these transmitters have been chosen to give good connection links for each receiver in the celling of the data center. Moreover, it is assumed that each ADT is composed of four branches with the direction of each branch being defined using two angles which are the Azimuth and Elevation angles. More specifically, the first ADT transmitter's azimuth angles of the four branches are set to 348º, 27º, 51º, and 63º, respectively, while their elevation angles are set to 20º, 18º, 13º, and 9º. Similarly, the second ADT transmitter's azimuth angles are set to 270º, 270º, 90º, and 90º, while the elevation angles are set to 18º, 45º, 45º, and 18º. Finally, the third ADT transmitter's azimuth angles are set to 270º, 270º, 270º, and 90º, while the elevation angles are set to 10º, 15.1º, 30º, and 73º.

Additionally, it is assumed that each branch's semi-angle at half-power is set to 2º similar to [26]. This setting will ensure a strong LOS component in the total received optical power in the 4 receivers at the ceiling. In other words, these values for azimuth angles, elevation angles, and the semi-angle at half-power of the ADT transmitters are selected to enable each branch to serve one receiver. It should be emphasized that the values for the azimuth angles and elevation angle of the ADT transmitters are calculated in a fashion similar to [44].

Furthermore, it is assumed that the three transmitters are pointed upwards, with each branch of each transmitter emitting an optical power of 150 mW [28], [29] as shown in Figs.1 and 2. Thus, it is assumed that the ToR switches produce four beams, with each beam directed at a spine switch.

*C. Optical Receiver Design*

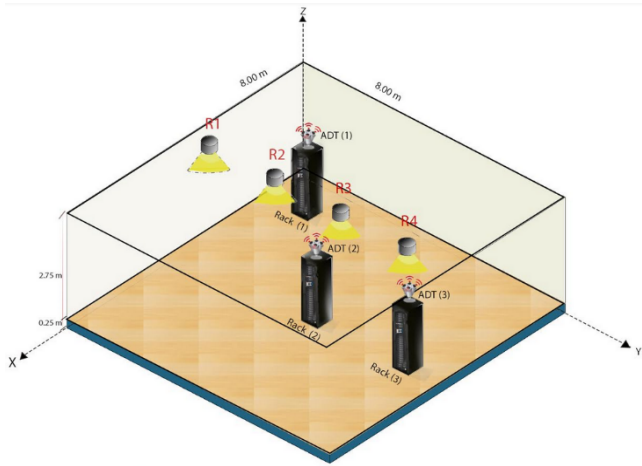

Figure 1 ADT with WFOV

The study utilizes two types of optical receivers. The first is WFOVR. The system has a row of optical receivers, which are placed in the middle of the ceiling, as shown in (Figs. 1 and 2). The goal of placing the receivers along the center of the data center room is to minimize the average values of the distance between the transmitters and the receivers to reduce the impact of the reflecting components and thus to ensure a strong LoS component between them. Their locations are (4m, 1m, 3m), (4m, 3m, 3m), (4m, 5m, 3m), and (4m, 7m, 3m). The current work uses a WFOVR (see Fig. 1) with an area of 20 $mm^2$ and FOV of 90º.

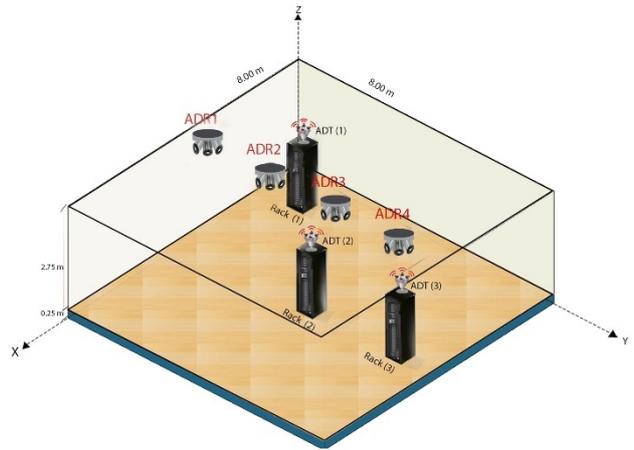

Figure 2 ADT with ADR

The second type of optical receivers is ADR (see Fig. 2) with a FOV of 5 º. The first receiver at 4m, 1m, 3m, has azimuth angles of 192.5º, 90º, and 90º and elevation angles of 19.5º, 18.5º, and 10º. The second receiver at 4m, 3m, 3m, has azimuth angles of 207.5º, 90º, and 90º and elevation angles of 18º, 45º, and 15º. The third receiver at 4m, 5m, 3m, has azimuth angles of 231.5º, 270º, and 90º and elevation angles of 13º, 45º, and 30.5º. The fourth receiver at 4m, 7m, 3m, has azimuth angles of 243.5º, 270º, and 270 º and elevation angles of 9.5º, 18.5º, and 73.5º. All receivers have an area equal to 20 $mm^2$ to support high data rates. Additionally, the optical receivers' responsivity of the silicon photodetector is assumed to be 0.6 A/W at near-infrared wavelengths of about 850 nm [45].

## IV. RESULTS

This section describes our evaluation of the proposed ADT systems to verify the performance of the IR links. More specifically, this method is utilized to analyze, study, and model indoor OWC channel features using a ray-tracing algorithm for direct LOS and first and second-order reflections [31]. The results of the ray tracing were generated using MATLAB and are presented using the corresponding data rate and SNR value.

Notably, when we evaluated the results while considering the first order reflections in our system with ADT, we found that the delay spread was negligible due to the very strong LoS component and weak first order reflections. This happened because the location of the receiver was fixed on the ceiling and a LoS component exists between transmitter and receiver and is not expected to be interrupted in the data centre setting between the ToR switch and the spine switch feed point on the ceiling. Furthermore, because the ADT has a narrow beam, limited first order reflections from the reach the receiver and hence this produced a flat channel due to the low reflection components.

The study involved the examination of two forms of optical receivers: WFOVRs and an ADR with three branches. We consider OOK modulation in which the probability of error $P_e$ is given as [20] and [31, 46]:

$$P_e = Q\sqrt{SNR}, \quad (1)$$

where Q(.) is the Gaussian function [47] and the electrical SNR is calculated as follows [48]:

$$SNR = \frac{R^2(P_{s1} - P_{s0})^2}{\sigma_t^2}, \quad (3)$$

where $R$ is photodetector responsivity, $P_{s1}$ is the optical power associated with logic 1, $P_{s0}$ is the optical power associated with logic 0, and $\sigma_t^2$ is the total noise due to the received signal and receiver's noise current spectral density. The receiver's input noise current was 4.47 pA/√Hz for the 5 GHz receiver used [25], [26]; $\sigma_t^2$ can be calculated as follows:

$$\sigma_t^2 = \sigma_{pr}^2 + \sigma_{bn}^2 + \sigma_{sig}^2 \quad (4)$$

where $\sigma_{pr}^2$ is the mean square receiver preamplifier noise, $\sigma_{bn}^2$ is the mean square background shot noise and $\sigma_{sig}^2$ is the mean square signal induced shot noise.

The achievable data rate is determined as [49]:

$$Channel\ capacity = B\ log_2(1 + SNR) \quad (5)$$

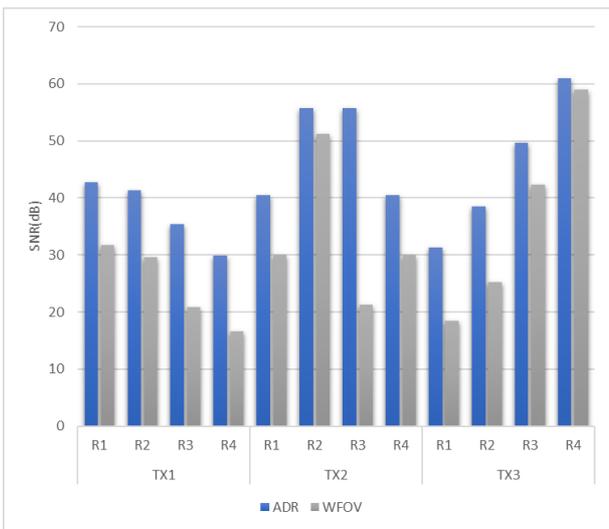

Figure 3 SNR results for the ADTs with ADR and WFOV receivers

where B is the bandwidth. Fig. 3 illustrates the SNR measured at each receiver (i.e. at receiver 1 (R1), at receiver 2 (R2), at receiver 3 (R3), and at receiver 4 (R4) from each transmitter (i.e. transmitter 1 (TX1), transmitter 2 (TX2), and transmitter 3 (TX3)). In comparing the SNR results for the ADR and WFOVR receivers, the results show that both receivers can provide a high SNR of 15.6 dB or higher, which is required for a $P_e$ of $10^{-9}$. However, the ADR provides a significantly higher SNR than WFOVR. The ADR and WFOVR offer data rates up to 19 Gbps as shown in Fig. 4. The WFOV receivers with 90º FOV collect a larger signal power, but admit more noise, while the ADR receivers with FOV 5º achieve higher SNR given the set up described wher the LoS signal component dominates and given the noise sources which are significant. It is shown that receiver (R1) has the lowest performance for transmitter 3 and that receiver (R4) has the lowest performance for transmitter 1. This is due to the fact that the distance between this receiver and its ADT is high since the rack is placed at the room edge while the receiver is placed near the midpoint in the x-direction. Hence, more power is lost along the path which in turn leads to a lower SNR. Additionally, it is shown that receivers R2 and R3 have the same performance for transmitter 2, while receivers R1 and R4 have the same performance for the same transmitter (ADT2). This is expected because of the equal distance from R2 and R3 to ADT2 as well as the equal distance from R1 and R4 to ADT2. Moreover, as expected, it is observed that the results for receiver (R4) are close to each other for both transmitters when using ADR and WFOVR.

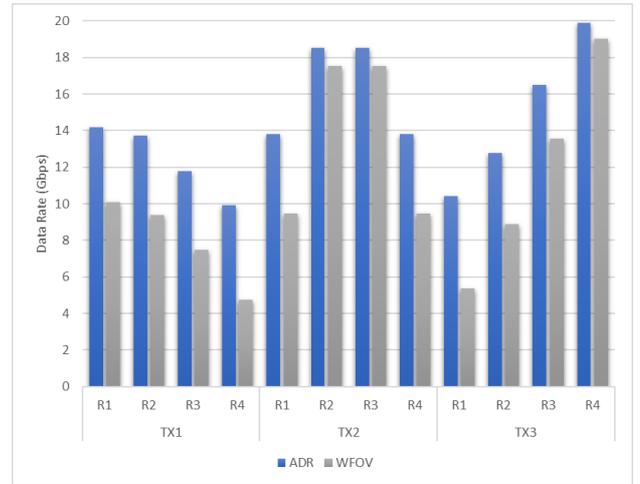

Figure 4 Achievable data rate when using ADR and WFOV receivers

This, is because it has the shortest distance to ADT3 given that it is placed directly above it. In a similar fashion, Fig. 3 illustrates that the rack placed at the edge of the data centre has a short link with the spine switch feed point, which is placed over it. Noting that a shorter communication distance leads to reduced power loss and, as a result, a higher SNR. The achievable data rate can be determined using (5). Fig. 4 presents the achievable data rate of our proposed uplink. Due to the long distance between the rack and receiver the SNR values are lower and accordingly lower data rates are achieved.

## V. CONCLUSIONS

This paper proposed the use of ADTs with ADR and WFOV receivers in the ceiling to achieve OWC uplinks to support connectivity between top of rack leaf switches and spine switches feed points in the ceiling in data centers. The ADTs and the two receiver types were designed and their SNR and data rate performances were evaluated. The proposed data center has three racks and four receivers. The racks have separate ADTs positioned at the top, while the receivers are mounted on the data center's room ceiling and linked to spine switches. Furthermore, each of the receivers serves a single branch of an ADT. The results demonstrate that the system can achieve an average data rate of 19 Gbps using simple OOK modulation while all the links have SNR value equal to or greater than 15.6 dB. For future work, we will consider implementing a spine-and-leaf data centre architecture using optical wireless links for the downlinks. We will also investigate the impact of realistic turbulent channels between the ToR switch and spine switch on the ceiling. Turbulence here may be caused by the hot air emitted by eth servers and the cold air used to achieve cooling in the data centre.

## ACKNOWLEDGMENTS

The authors would like to acknowledge funding from the Engineering and Physical Sciences Research Council (EPSRC) INTERNET (EP/H040536/1), STAR (EP/K016873/1) and TOWS (EP/S016570/1) projects. The authors extend their appreciation to the deanship of Scientific Research under the International Scientific Partnership. ASA would like to thank Taibah University in the Kingdom of Saudi Arabia for funding her PhD scholarship. All data are provided in full in the results section of this paper.